\documentclass[12pt,titlepage]{article}
\renewcommand{\theequation}{\thesection.\arabic{equation}}
\renewcommand{\to}{\rightarrow}

\newcommand{\beq}{\begin{equation}}
\newcommand{\eeq}{\end{equation}}
\newcommand{\bea}{\begin{eqnarray}}
\newcommand{\eea}{\end{eqnarray}}
\newcommand{\eq}{\begin{equation}}
\newcommand{\en}{\end{equation}}
\newcommand{\eqa}{\begin{eqnarray}}
\newcommand{\ena}{\end{eqnarray}}

\def\ln{{\rm ln}}

\begin{document}
\thispagestyle{empty}
\begin{titlepage}
\addtolength{\baselineskip}{.7mm}
\thispagestyle{empty}
\begin{flushright}
DFTT 19/05\\
\end{flushright}
\vspace{7mm}
\begin{center}
{\large 
{\bf Symmetric space description of carbon nanotubes 
}}\\[11mm]
{
\bf 
Michele~Caselle$^1$ and Ulrika~Magnea$^2$ 
} \\
\vspace{3mm}
{\it Department of Theoretical Physics, University of Torino \\
and INFN, Sez. di Torino\\
Via P. Giuria 1, I-10125 Torino, Italy \\ 
{\ }\\
$^1${\tt caselle@to.infn.it} \\
$^2${\tt blom@to.infn.it} 
\\[4mm]
\vspace{15mm}}

{\bf Abstract}\\[5mm]
\end{center}

Using an innovative technique arising from the theory of symmetric
spaces, we obtain an approximate analytic solution of the
Dorokhov--Mello--Pereyra--Kumar (DMPK) equation in the insulating
regime of a metallic carbon nanotube with symplectic symmetry and an
odd number of conducting channels.  This symmetry class is
characterized by the presence of a perfectly conducting channel in the
limit of infinite length of the nanotube. The derivation of the DMPK
equation for this system has recently been performed by Takane, who
also obtained the average conductance both analytically and
numerically. Using the Jacobian corresponding to the transformation to
radial coordinates and the parameterization of the transfer matrix
given by Takane, we identify the ensemble of transfer matrices as the
symmetric space of negative curvature $SO^*(4m+2)/[SU(2m+1)\times
U(1)]$ belonging to the $DIII$--odd Cartan class.  We rederive the
leading--order correction to the conductance of the perfectly
conducting channel $\langle {\rm ln}\,\delta g \rangle $ and its
variance ${\rm Var}({\rm ln}\, \delta g)$.  Our results are in
complete agreement with Takane's. In addition, our approach based on
the mapping to a symmetric space enables us to obtain new universal
quantities: a universal group theoretical expression for the ratio
${\rm Var}({\rm ln}\, \delta g)/\langle {\rm ln}\,\delta g\rangle $
and as a byproduct, a novel expression for the localization length for
the most general case of a symmetric space with $BC_m$ root system, in
which all three types of roots are present.

\end{titlepage}

\newpage
\setcounter{footnote}{0}

\section{Introduction}
\label{sec-intro}

Carbon nanotubes, consisting of single or multiple graphene layers
wrapped into nanometer--thick cylinders\footnote{Stacked sheets of
single graphene layers form graphite.}, have been an object of
experimental and theoretical study in mesoscopic physics over the past
decade.  They exhibit particular transport properties due to the
special characteristics that distinguish them from ordinary quasi--one
dimensional quantum wires. The hexagonal lattice of carbon atoms puts
constraints on the possibilities of matching a single layer of atoms
onto itself to form a cylinder or torus, giving rise to systems with
differing discrete chirality, and in the case of a toroidal geometry,
differing degree of twist. The cylindrical or toroidal geometry allows
the conduction electrons to orbit around the axis of the
cylinder. This happens for certain chiralities only.  Experimental and
theoretical analyses show that carbon nanotubes are either metallic or
semiconducting. The electrical properties of carbon nanotubes are
governed by electrons close to the Fermi energy and show some
interesting phenomenology (see e.g. \cite{SasaKawa}).

A carbon nanotube can be considered a special kind of disordered
quasi--one dimensional quantum wire, and the same techniques that are
applied in the theoretical study of mesoscopic wires can be applied
also to carbon nanotubes.  Disordered systems were classified by Dyson
into three standard symmetry classes (orthogonal, unitary and
symplectic, labelled by Dyson index $\beta=1,\ 2$, and $4$
respectively) \cite{Dyson}.  In this paper we will focus on metallic
carbon nanotubes with symplectic symmetry.  By this we mean that the
system is time--reversal invariant but has strong spin--orbit
scattering, according to Dyson's classification. 

Such nanowires have been studied by Ando and Suzuura \cite{AS}, who 
showed that they have an anomalous transport property: there is one 
perfectly conducting channel present even in the limit of an infinitely 
long wire. The first discovery of the non--vanishing conductance in the
long--wire limit in the symplectic symmetry class is, however, due to
Zirnbauer \cite{Z} and to Mirlin, M\"uller--Groeling, and Zirnbauer
\cite{MMZ}. These authors expressed moments of the conductance of thick
disordered wires in terms of the heat kernel in a supersymmetric
$\sigma$--model approach, subsequently obtaining approximate analytical 
expressions for the conductance for all three symmetry classes by Fourier 
analysis on the supersymmetric manifold.

Takane \cite{T1,T2,T3} and Sakai and Takane \cite{ST} conducted
further studies of systems with symplectic symmetry and with an even or odd
number of conducting channels, employing a variety of methods.  It was
confirmed in \cite{T1} and \cite{T2}, using the supersymmetric formalism
and numerical simulations, respectively, that for an {\it odd} number
of conducting channels (which is the case in a carbon nanotube), the
dimensionless conductance $g\to 1$ in the limit of an infinitely long
wire, indicating the absence of Anderson localization.  In this case
there is an unpaired transmission eigenvalue equal to unity which
gives rise to the anomalous behavior of the conductance.  This is to
be contrasted with the case of an {\it even} number of conducting
channels, when the average dimensionless conductance approaches zero
with increasing wire length $L$ (this case is realized in an ordinary
quantum wire, where the limit $L\to \infty $ corresponds to the
insulating regime).

In \cite{T3} these results were confirmed using scaling theory. Within
the framework of random matrix theory, the
Dorokhov--Mello--Pereyra--Kumar (DMPK) equation for the probability
distribution of the transmission eigenvalues as a function of the
length of the wire was constructed for this particular system, and the
conductance and its variance were derived in the limit of an
infinitely long wire.  In contrast, it was proved in \cite{ST} that
the presence of the perfectly conducting channel in the odd--channel
case does not have any bearing on the conductance and its variance if
the length of the wire is shorter than the localization length, in
which case there is almost no even--odd difference. This is due to
geometric eigenvalue repulsion in random matrix theory. As a
consequence of the presence of the unpaired transmission
eigenvalue\footnote{We use Takane's notation in which $N=2m$ in the
even--channel case and $N=2m+1$ in the odd--channel case. In this
notation, the number of non--zero, {\it distinct}, paired eigenvalues
in the symplectic symmetry class is $m$. (This quantity is often
denoted by $N$ elsewhere.)}  $T_{2m+1}=1$, the other eigenvalues are
repelled and therefore diminish (we recall that $0\leq T_i \leq 1$).

This paper is organized as follows.  Sections \ref{sec-SS} and
\ref{sec-scatt} contain some generalities concerning the symmetric
space description of random matrix theories and the standard
theoretical description of mesoscopic scattering, respectively.  In
section~\ref{sec-diff} we discuss the mapping of the DMPK equation to
the equation of free diffusion on the symmetric space.  

In section \ref{sec-detss} we then identify the symmetric space
corresponding to the ensemble of transfer matrices for the symplectic
ensemble with an odd number of scattering channels.  This will be done
in two ways: first using the Jacobian on the symmetric space, and then
using the parametrization of the transfer matrix given by Takane
\cite{T3}.  In section \ref{sec-cond} we show how to solve the DMPK
equation and obtain the conductance of the system using the technique
of zonal spherical functions on the symmetric space. Some details of
the calculation are reported in Appendix \ref{app}.

\section{The symmetric space point of view}
\label{sec-SS}
\setcounter{equation}{0}

The classification of random matrix ensembles in terms of symmetric
coset spaces of Lie groups (or as subspaces of Lie algebras) is by now
rather well--known. In reference \cite{ss} a detailed review of this
subject was given. It was shown that all the known hermitean random
matrix ensembles can be identified with symmetric coset spaces of Lie
groups or with subspaces of Lie algebras. There are exactly
twelve\footnote{In Cartan's classification there were eleven but it is
natural to split the $DIII$--symmetry class into even and odd,
following the example of some recent authors. This is because the even
and odd cases correspond to different restricted root systems.}  such
groups of symmetric spaces.  Each group comprises three symmetric
spaces identified with manifolds of positive, zero, and negative
curvature, respectively.  The positive curvature spaces are identified
with ensembles of scattering matrices, the zero curvature spaces
(which are Lie algebra subspaces) with ensembles of Hamiltonians, and
the negative curvature spaces with ensembles of transfer matrices.  

In the third and fourth column of Table~\ref{tab3}, which we have
reproduced from reference \cite{ss}, the symmetric coset spaces of
positive and negative curvature arising from simple Lie groups are
listed (the zero curvature spaces are not listed since these are
simply identified with the Lie algebra subspace defining the
corresponding curved manifolds). In the fifth through seventh column
of Table~\ref{tab3} the multiplicities of the roots of the
corresponding restricted root systems are listed. The subscripts refer
to ordinary, long and short roots, respectively.

\begin{table}
\caption{Cartan's classification of irreducible symmetric spaces and
some of their random matrix theory realizations. In the third and
fourth column we have listed the symmetric coset spaces of positive
and negative curvature based on simple Lie groups. The zero curvature
spaces are not listed since these are simply identified with the Lie
algebra subspace defining the corresponding curved manifolds. $m_o$,
$m_l$, and $m_s$ are the multiplicities for the respective restricted
roots and in the last three columns we find the random matrix
ensembles with known physical applications corresponding to symmetric
spaces of positive, zero and negative curvature, respectively.  The
ensemble of transfer matrices for symplectic odd--channel carbon
nanotubes corresponding to the present case has been inserted in
boldface in the last column. We have set $\nu \equiv p-q$ and for the
random matrix ensembles we use the following abbreviations: C for
circular, G for gaussian, $\chi $ for chiral, B for Bogoliubov--de
Gennes, P for $p$--wave, T for transfer matrix and S for S--matrix
ensembles.  The upper indices indicate the curvature of the respective
symmetric spaces, while the lower indices correspond to the
multiplicities of the restricted roots characterizing each
triplet.\label{tab3}} \vskip10mm

\hskip-1cm
\begin{tabular}{|l|l|l|l|l|l|l|l|l|l|}
\hline
$\begin{array}{c}Restricted\\ root\ space\end{array}$ & $\begin{array}{c}Cartan\\ class \end{array}$ & $G/K\ (G)$ & $G^*/K\ (G^C/G)$ & $m_o$ & $m_l$ & $m_s$ & $X^+$ & $X^0$ & $X^-$\\
\hline

$A_{N-1}$     & A    & $SU(N)$                 & $\frac{SL(N,C)}{SU(N)}$ & 2 & 0 & 0 & C$^+_{2,0,0}$ & G$^0_{2,0,0}$ & ${\rm T}^-_{2,0,0}$ \\ 
$A_{N-1}$     & AI   & $\frac{SU(N)}{SO(N)}$   & $\frac{SL(N,R)}{SO(N)}$ & 1 & 0 & 0 & C$^+_{1,0,0}$ & G$^0_{1,0,0}$ & ${\rm T}^-_{1,0,0}$\\ 
$A_{N-1}$     & AII  & $\frac{SU(2N)}{USp(2N)}$ & $\frac{SU^*(2N)}{USp(2N)}$ & 4 & 0 & 0 & C$^+_{4,0,0}$&G$^0_{4,0,0}$& ${\rm T}^-_{4,0,0}$\\
\hskip-2mm $\begin{array}{l} BC_q\ {\scriptstyle (p>q)} \\ C_q\  {\scriptstyle (p=q)} \end{array}$ 
                          & AIII & $\frac{SU(p+q)}{SU(p)\times SU(q)\times U(1)}$ & $\frac{SU(p,q)}{SU(p)\times SU(q)\times U(1)}$ & 2 & 1 & $2\nu $  & \hskip-2mm $\begin{array}{l}  \\ {\rm S}^+_{2,1,0}\end{array}$  & 
 $\chi^0_{2,1,2\nu }$ & 
 \hskip-2mm $\begin{array}{l}  \\ {\rm T}^-_{2,1,0} \end{array}$  \\
\hline

$B_N$          & B    & $SO(2N+1) $               & $\frac{SO(2N+1,C)}{SO(2N+1)}$ & 2 & 0 & 2  &   & P$^0_{2,0,2}$ &  \\ 

\hline

$C_N$ & C    & $USp(2N)$                               &$\frac{Sp(2N,C)}{USp(2N)}$ & 2 & 2 & 0 & B$^+_{2,2,0}$& B$^0_{2,2,0}$ & ${\rm T}^-_{2,2,0}$\\
$C_N$ & CI   & $\frac{USp(2N)}{SU(N)\times U(1)}$      & $\frac{Sp(2N,R)}{SU(N)\times U(1)}$& 1 & 1 & 0  & B$^+_{1,1,0}$ & B$^0_{1,1,0}$  & T$^-_{1,1,0}$  \\
\hskip-2mm $\begin{array}{l} BC_q\  {\scriptstyle (p>q)} \\  C_q\  {\scriptstyle (p=q)} \end{array}$ 
                  & CII & $\frac{USp(2p+2q)}{USp(2p)\times USp(2q)}$ & $\frac{USp(2p,2q)}{USp(2p)\times USp(2q)}$& 4 & 3 & $4\nu $ &   & $\chi^0_{4,3,4\nu }$ & \hskip-2mm $\begin{array}{l} \\ {\rm T}^-_{4,3,0}\end{array}$   \\
\hline

$D_N$    & D          &   $SO(2N)$                            & $\frac{SO(2N,C)}{SO(2N)}$ & 2 & 0 & 0 & B$^+_{2,0,0}$  &  B$^0_{2,0,0}$  & ${\rm T}^-_{2,0,0}$\\
$C_N$    & DIII-e  & $\frac{SO(4N)}{SU(2N)\times U(1)}$     & $\frac{SO^*(4N)}{SU(2N)\times U(1)}$ & 4 & 1 & 0 & B$^+_{4,1,0}$ & B$^0_{4,1,0}$ & T$^-_{4,1,0}$  \\
$BC_N$   & DIII-o  & $\frac{SO(4N+2)}{SU(2N+1)\times U(1)}$ & $\frac{SO^*(4N+2)}{SU(2N+1)\times U(1)}$& 4 & 1 & 4 &  & P$^0_{4,1,4}$ & {\bf T$^-_{4,1,4}$}   \\
\hline

\hskip-2mm $\begin{array}{l} B_q\ {\scriptstyle (p>q)}\\ D_q\ {\scriptstyle (p=q)} \end{array}$
            & BDI & $\frac{SO(p+q)}{SO(p)\times SO(q)}$   & $\frac{SO(p,q)}{SO(p)\times SO(q)}$ & 1 & 0 & $\nu $ &   & $\chi^0_{1,0,\nu }$ & $\begin{array}{l} \\ {\rm T}^-_{1,0,0} \end{array}$\\

\hline
\end{tabular}
\end{table}

Let us remind the reader of the construction of the root system of a
simple Lie algebra. There are two kinds of generators in the algebra:
there is a maximal abelian subalgebra $\{H_1,...,H_r\}$, where $r$ is
the rank of the algebra and $[H_i,H_j]=0$, and there are raising and
lowering operators $E_{\pm\alpha}$. The roots $\alpha $ are functionals
on the Cartan subalgebra whose components $\alpha_i $ satisfy

\beq
[H_i,E_\alpha]=\alpha_iE_\alpha
\eeq

The angle between root vectors can take only a few discrete values,
which makes Cartan's classification possible and leads, for the
non--exceptional Lie groups, to four infinite series of root systems
$A_{n-1},\ B_n,\ C_n,$ and $D_n$.  Just as a root system corresponds
to each simple Lie group, for each symmetric coset space one can
identify a unique {\it restricted} root system where each root vector
has a given multiplicity.  It would bring us too far to discuss the
construction of these root systems here, and we refer the interested
reader to \cite{ss} for details. In any case it is similar to the
construction of the root system for the entire Lie algebra. 

The non--reduced 
root system $BC_n$ is defined as the union of $B_n$ and $C_n$.  
It is of relevance in this context as it appears as the restricted root 
system of several types of symmetric spaces, as can be seen by 
inspection of Table~\ref{tab3}.  

In this framework, the multiplicities of restricted root vectors
associated to a symmetric coset space are identified in the random
matrix ensemble as the exponents appearing in the Jacobian for the
transformation from the space of random matrices to the space of
random matrix eigenvalues. This Jacobian gives rise to the
characteristic eigenvalue repulsion in random matrix theory. Such a
transformation may be expressed as

\beq
S\to U^{-1}X U
\eeq

where $S$ is a random matrix, $X$ is (block--)diagonal and contains
the random matrix eigenvalues and $U$ is unitary. The random matrix
partition function is invariant under such a transformation.  For
example, for the simple gaussian unitary random matrix ensemble
characterized by root multiplicities $m_o=2$, $m_l=m_s=0$, the
Jacobian of such a coordinate transformation is given by
$J(\{x_i\})\sim \prod_{i<j} (x_i-x_j)^2$ (where $\{x_i\}$ are the
eigenvalues). In this case there is only one exponent in the Jacobian,
$\beta \equiv 2$ (where $\beta $ indeed is Dyson's symmetry index)
which is given by the multiplicity of ordinary roots: $\beta = m_o =
2$. Similar relations between the exponents in the Jacobian and the
three root multiplicities are true also in the case of more
complicated Jacobians with more than one exponent. We conclude that
the root structure of the underlying symmetric space completely
determines the celebrated universal random matrix eigenvalue
correlations!

The random matrix ensembles identified with the symmetric spaces in
each row are listed in the last three columns in Table~\ref{tab3}. The
notation for the various types of ensemble is as follows: C for
circular, G for gaussian, $\chi $ for chiral, B for Bogoliubov--de
Gennes, P for $p$--wave, T for transfer matrix and S for S--matrix
ensembles. A discussion of these ensembles here is outside the scope
of this paper and we refer to \cite{ss} or to the vast random matrix
literature for an introduction.

\section{Scattering in a mesoscopic wire}
\label{sec-scatt}
\setcounter{equation}{0}

All the material in this section is standard and is included only to make
the paper self--contained.

The scattering of electrons in a mesoscopic wire attached to ideal
leads can be described using the transfer matrix.  Assuming there are
$m$ propagating modes at the Fermi level, we describe them by a vector
of length $2m$ of incoming modes $I$, $I'$ and a similar vector of
outgoing modes $O$, $O'$ in each lead. Let the unprimed letters denote
the modes in the left lead and the primed letters the modes in the
right lead. While the scattering matrix $S$ relates the incoming wave
amplitudes $I$, $I'$ to the outgoing wave amplitudes $O$, $O'$, the
transfer matrix $M$ relates the wave amplitudes in the left lead to
those in the right lead:

\beq
M\left( \begin{array}{c} I \\ O \end{array} \right) =
\left( \begin{array}{c} O' \\ I' \end{array} \right)
\eeq

The transmission eigenvalues $T_i$ are the $m$ eigenvalues of the
matrices $tt^\dagger$ (or equivalently, $t't'^\dagger$), where $t$
($t'$) is the left--to--right (right--to--left) transmission matrix
appearing in the scattering matrix (and $r$, $r'$ are the respective
reflection matrices):

\beq
S=\left( \begin{array}{cc} r & t' \\
                          t & r' \end{array} \right)
\eeq

The transfer matrix is expressed in terms of these submatrices as
\cite{MelloTom}

\beq
M=\left( \begin{array}{cc} (t^\dagger)^{-1} & r't'^{-1} \\
                           -t'^{-1}r & t'^{-1} \end{array} \right)
\eeq

In terms of the transmission eigenvalues $\{T_i\}$, the expression for
the dimensionless conductance is given in the Landauer--Lee--Fisher
theory by

\beq
\label{eq:g}
g=\frac {G}{G_0}=\sum_{i=1}^m T_i\ \ \ \ \ \ \ (G_0=\frac {2e^2}{h})
\eeq

Instead of the transmission eigenvalues, it is common to use the
non--negative variables $\{\lambda_i\}$ defined by

\beq
\label{eq:Tlambda}
\lambda_i=\frac{1-T_i}{T_i}
\eeq
 
$M$ can then be parametrized as\footnote{This parametrization is valid
for the cases $M\in Sp(2m,R)/[SU(m)\times U(1)]$ (``orthogonal'' ensemble) and $M\in
SO^*(4m)/[SU(2m)\times U(1)]$ (``symplectic'' ensemble), whereas for the unitary ensemble
$SU(m,m)/[SU(m)\times SU(m)\times U(1)]$, $u^*$ and $v^*$ have to be
substituted by two unitary matrices $u'$ and $v'$ not related to $u$
and $v$ by complex conjugation.} \cite{MPK}

\beq
\label{eq:Mpara}
M=\left( \begin{array}{cc} u & 0 \\
                           0 & u^* \end{array} \right) 
\left( \begin{array}{cc} \sqrt{1+\Lambda} & \sqrt{\Lambda} \\
                         \sqrt{\Lambda} & \sqrt{1+\Lambda}\end{array} \right) 
\left( \begin{array}{cc} v & 0 \\
                         0 & v^* \end{array} \right) \equiv U\Gamma V
\eeq

where $u$, $v$ are unitary $m \times m$ matrices and 

\beq
\label{eq:Lambda}
\Lambda={\rm diag}(\lambda_1,...,\lambda_m) 
\eeq

is a diagonal matrix. 

In the symplectic (even--channel) case, there is a {\it doubling} of
the degrees of freedom due to the fact that the components of $I$,
$O$, etc. become {\it spinors}. This means the eigenvalues
$\{\lambda_i\}$, $i=1,...,m$ are doubled so that we get $m$ pairs of
degenerate eigenvalues (this is referred to as Kramers degeneracy).
If we write $u$, $v$ and $\Lambda $ in terms of {\it complex}
matrices, we are dealing with $2m\times 2m$ matrices, whereas in terms
of {\it quaternion} matrices, they are $m\times m$. The number of {\it
distinct} eigenvalues is $m$. In this case eq.~(\ref{eq:Mpara}) and
(\ref{eq:Lambda}) are still valid, but the matrix elements are now
quaternions. Following Takane, in the even--channel case we then set
$N=2m$ and in the odd--channel case $N=2m+1$, so that the matrix $M$
in the symplectic ensemble becomes a matrix with $N\times N$ complex
elements.

It is known that the parametrization (\ref{eq:Mpara}) leads to a coset
space structure for $N=2m$ \cite{h90,MC}. One can show that the
symmetries -- in this case flux conservation, time reversal symmetry,
and no spin--rotation invariance -- imposed on $M$ lead to $M\in
SO^*(4m)$ for the symplectic even--channel case, in which all the
eigenvalues are paired due to Kramers degeneracy and there is no zero
eigenvalue. It is easy to see that the eigenvalues $\{\lambda_i\}$ are
unchanged if one transforms the matrix $M$ in the following way:

\beq
\label{eq:rotM}
M\to M'=WU\Gamma VW^{-1}=U'\Gamma V'
\eeq

where $W$ for the symplectic ensemble is a unitary quaternion
matrix. Thus, in the symplectic even--channel case, $M$ belongs to the
symmetric coset space $SO^*(4m)/[SU(2m)\times U(1)]$ corresponding to
the $DIII$--even Cartan class. 

\section{DMPK equation as free diffusion on a symmetric space}
\label{sec-diff}
\setcounter{equation}{0}

The DMPK equation is a differential equation describing the evolution
of the transmission eigenvalues of a quantum wire with increasing
length of the wire.  Its solution gives access to the distribution of
transmission eigenvalues and therefore to the conductance through
equation~(\ref{eq:g}).

In the random matrix approach to quantum wires, as a consequence of
the description of random (transfer) matrix ensembles in terms of
symmetric spaces, the DMPK equation of a quantum wire can be expressed
in terms of the radial part of the Laplace--Beltrami operator on the
appropriate symmetric space.  The Laplace--Beltrami operator (denoted
$\Delta_B$) is simply the operator of free diffusion on a given
symmetric space, and corresponds to the Laplacian in ${\bf R^3}$.  It
is defined as the lowest order Casimir operator of the algebra,
expressed in local coordinates as a differential operator. We denote
its radial part by $\Delta_B'$. As an example, the Laplace--Beltrami
operator on the symmetric space $SO(3)/SO(2)$ (whose points are in
one--to--one correspondence with the two--sphere) is given, in radial
coordinates, by $\Delta_B=\partial_\theta^2+{\rm cot}\theta\,
\partial_\theta+ {\rm sin}^{-2}\theta \, \partial_\phi^2$ (we have set
the radius of the sphere $r\equiv 1$), and its radial part is
$\Delta_B'=\partial_\theta^2+{\rm cot}\theta\, \partial_\theta $.

It is a completely general result, discussed for example in 
\cite{c1,ss}, that the DMPK equation for the transmission eigenvalues
can be written as

\begin{equation}
\label{eq:DMPKbis}
\frac{\partial P(\{x_i\},s)}{\partial s}=
(2\gamma )^{-1}BP(\{x_i\},s)
\end{equation}

where $\gamma $ is a constant and $\{x_i\}$ are radial coordinates on
the symmetric space. The operator

\begin{equation}
B=\sum_{k=1}^m\frac{\partial}{\partial x_k}J(\{x_i\})
\frac{\partial}{\partial x_k} J^{-1}(\{x_i\})
\label{defb}
\end{equation}

is related to the radial part of the Laplace--Beltrami operator
$\Delta_B'$ on the symmetric space through 

\beq
\label{eq:JBJ}
\Delta_B'=J^{-1}BJ
\eeq

and $J(\{x_i\})$ is the Jacobian for the transformation from the space of
random matrices to the space of eigenvalues (radial coordinates),
cf. section~\ref{sec-SS}. The constant $\gamma $ is in the most
general case given by (cf. the expression in reference~\cite{MC} where
$\theta =0$)

\beq
\label{eq:gamm}
\gamma = \frac{2(1+\eta +\beta (m-1)+\theta /2)}{1+\eta }  
\eeq

where $\eta \equiv m_l$, $\beta \equiv m_o$, $\theta \equiv m_s$ and
$m_l,\ m_o,\ m_s$ denote the multiplicities of the long, ordinary, and
short (restricted) roots of the symmetric space and $\beta $
is Dyson's index. This expression was obtained simply by matching 
$\gamma $ to the explicitly known cases. In eq.~(\ref{eq:DMPKbis}) 
the dimensionless variable $s$ is defined by $s=L/l$, where $L$ is the 
length of the mesoscopic wire and $l$ is the mean free path of the 
scattered electrons.

\section{Determination of the symmetric space}
\label{sec-detss}
\setcounter{equation}{0}

In this section we will determine the symmetric space to which the
transfer matrix of the symplectic odd--channel carbon nanotube  
belongs, using two different methods.

\subsection{Determination using the Jacobian}

The DMPK equation for the symplectic ensemble with an odd number
$N=2m+1$ of conducting channels was derived in \cite{T3} and found to
be given by

\beq
\label{eq:DMPK}
\frac {\partial P(\lambda_1,...,\lambda_m,s)}{\partial s}=
\frac{1}{N-1} \sum_{i=1}^m \frac{\partial}{\partial \lambda_i}
\left( \lambda_i (1+\lambda_i )\tilde{J} \frac{\partial}{\partial \lambda_i} 
\frac {P(\lambda_1,...,\lambda_m,s)}{\tilde{J}} \right ) 
\eeq

where we believe there is a typing mistake in a sign in the formula
given in \cite{T3}. Note that there are $m$ pairs of distinct
non--zero eigenvalues and one zero eigenvalue in this case. For an odd
number of conducting channels, $\tilde{J}$ is given by

\beq
\label{eq:jacobian-o}
\tilde{J}_o(\{\lambda_i\})=\prod_{i=1}^m \lambda_i^2 \prod_{j>k}^m (\lambda_j - \lambda_k)^4
\eeq

In \cite{ST}, the corresponding function was given for the even--channel case:

\beq
\label{eq:jacobian-e}
\tilde{J}_e(\{\lambda_i\})=\prod_{j>k}^m (\lambda_j - \lambda_k)^4
\eeq

The coordinates $\{\lambda_i\}$ are not the radial coordinates on the
symmetric space. By performing the variable substitution

\beq
\label{eq:subs}
\lambda_i={\rm sinh}^2x_i
\eeq

in the {\it integration measure}\footnote{It is important to note that
$\tilde{J}_o(\{\lambda_i\})\prod_i {\rm d}\lambda_i$ is the {\it integration
measure} on the symmetric space expressed in the ``wrong''
coordinates. Simply doing the substitution (\ref{eq:subs}) in
$\tilde{J}_o(\{\lambda_i\})$, we would miss the factor coming from the
long roots. This is evident in going from eq.~(\ref{eq:DMPKbis}) to
eq.~(\ref{eq:DMPK}), where we have to take into account also the
factors of $\frac{\partial \lambda_j}{\partial x_i}$ in the chain rule
expression for $\frac{\partial }{\partial x_i}$.}

\beq
\tilde{J}_o(\{\lambda_i\})\prod_i{\rm d}\lambda_i   \to J_o(\{x_i\})\prod_i {\rm d}x_i
\eeq 

and using identitites for the hyperbolic functions, we obtain the
Jacobian on the symmetric space, expressed in the radial coordinates $\{x_i\}$:

\beq
\label{eq:jacobianox}
J_o(\{x_i\})=\prod_{i=1}^m {\rm sinh}^4 x_i \prod_{j>k}^m {\rm sinh}^4(x_j-x_k) 
{\rm sinh}^4(x_j+x_k) \prod_{l=1}^m {\rm sinh}(2x_l)
\eeq

Using now the general result for the Jacobian of the transformation to
radial coordinates on a symmetric space, discussed in detail in
\cite{ss},

\beq
\label{eq:J_j}
\begin{array}{l}
J^{(0)}(x)=\prod_{\alpha \in R^+} (x^\alpha )^{m_\alpha }\\
\\
J^{(-)}(x)=\prod_{\alpha \in R^+} ({\rm sinh}(x^\alpha ))^{m_\alpha }\\ 
\\
J^{(+)}(x)=\prod_{\alpha \in R^+} ({\rm sin}(x^\alpha ))^{m_\alpha }\end{array}
\eeq

where $0,-,+$ denote the curvature of the space, $R^+$ denotes the set
of positive restricted roots, $x^\alpha $ denotes the projection of
$x$ on the root $\alpha $, and $m_\alpha $ its multiplicity, we
conclude that eq.~(\ref{eq:jacobianox}) describes the Jacobian on a
symmetric space with negative curvature and root multiplicities
$m_s=4,\ m_o=4,\ m_l=1$ for the short, ordinary, and long roots,
respectively. Comparing with Table~\ref{tab3} we
conclude that the transfer matrix $M$ in the odd--channel case belongs
to the symmetric space $SO^*(4m+2)/[SU(2m+1)\times U(1)]$ (Cartan
class $DIII$--odd). We have inserted this ensemble in the last column
of Table~\ref{tab3} in boldface. 

A similar analysis of the Jacobian for the even--channel case using
eq.~(\ref{eq:jacobian-e}) results in the coset space
$SO^*(4m)/[SU(2m)\times U(1)]$ ($DIII$--even), which agrees with the
known result from the parametrization (\ref{eq:Mpara}) given above.

\subsection{Determination using a parametrization of $M$}

The symmetric coset space of $M$ in the odd--channel case can also be
identified using the parametrization

\beq
\label{eq:Tpara}
M_o=\left(\begin{array}{cc} {\rm e}^\theta & 0 \\ 0 & {\rm e}^{\theta^*} \end{array} \right)
\left(\begin{array}{cc} (1+\eta \eta^\dagger)^\frac{1}{2} & \eta \\
                         \eta^\dagger &   (1+\eta^\dagger \eta)^\frac{1}{2} \end{array} \right)
\eeq

given by Takane \cite{T3}. Here $\theta =ih$ where $h$ is an
$(2m+1)\times (2m+1)$ hermitean matrix and $\eta $ is an
$(2m+1)\times (2m+1)$ arbitrary complex {\it antisymmetric} matrix.

Let us compare this with the even--channel case. In \cite{MelloTom} it
was observed that in the even--channel case, the parametrization
(\ref{eq:Mpara}) is equivalent to

\beq
\label{eq:mellotompara}
M_e=\left(\begin{array}{cc} {\rm e}^\theta & 0 \\ 0 & {\rm e}^{\theta^*} \end{array} \right)
\left(\begin{array}{cc} (1+\zeta \zeta^*)^\frac{1}{2} & \zeta \\
                         \zeta^* &   (1+\zeta^* \zeta)^\frac{1}{2} \end{array} \right)
\eeq

with $\theta $ defined as above except that it is now a $2m\times 2m$
matrix, and $\zeta$ is an arbitrary $2m\times 2m$ complex {\it
symmetric} matrix, if one makes the identifications

\beq
\label{eq:id1}
{\rm e}^\theta \equiv uv \ \ \ \ \ \zeta \equiv v^\dagger \sqrt{\Lambda }v^*
\eeq
 
Note that this agrees with the properties of $\theta $ and
$\zeta$, if $\sqrt {\Lambda }$ is symmetric:

\beq
\sqrt {\Lambda }=\left(\begin{array}{ccc} \sqrt{\lambda_1}   &        & \\
                                                            & \ddots & \\
                                                            &        & \sqrt{\lambda_m}
\end{array} \right)
\eeq

where each eigenvalue is twofold degenerate.  It is easy to verify
that Takane's parametrization (\ref{eq:Tpara}) in the odd--channel
case is equivalent to (\ref{eq:Mpara}) if instead of equation
(\ref{eq:id1}) we identify

\beq
\label{eq:id2}
{\rm e}^\theta \equiv uv \ \ \ \ \ \eta \equiv v^\dagger \sqrt{\Lambda'}v^*
\eeq
 
where $u,\ v$ are unitary and $\sqrt{\Lambda'}$ denotes an {\it
antisymmetric} $(2m+1)\times (2m+1)$ matrix of the form

\beq
\label{eq:'}
\sqrt{\Lambda'}=\left(\begin{array}{ccccc} 
                               0                 & \sqrt{\lambda_1} &                      &                     &   \\
                               -\sqrt{\lambda_1} &  0               &                      &                     &   \\
                                                 &    \ddots        &                      &                     &   \\
                                                 &                  & 0                    & \sqrt{\lambda_{m}} &   \\
                                                 &                  & -\sqrt{\lambda_{m}} &  0                  &   \\
                                                 &                  &                      &                     & 0 \\
\end{array}\right)
\eeq

Note that {\it any} antisymmetric matrix $\eta $ can be written in the
form expressed in equations~(\ref{eq:id2},\ref{eq:'}).  Therefore we
can use the same argument as in the even--channel case
(eq.~(\ref{eq:rotM})) to derive the coset structure
$SO^*(4m+2)/[SU(2m+1)\times U(1)]$ of the ensemble of transfer
matrices in the odd--channel case. The restricted root system corresponding
to this symmetric space is of the type $BC_m$. This will be used in the
following section, where we will also remind the reader of some general
features of root systems. 

To our knowledge this is the first known physical realisation of this
ensemble, which in the notation of table~\ref{tab3} we denote
$T^-_{4,1,4}$. Thus we have now filled out the previously empty space
in Table~\ref{tab3} corresponding to this physical realisation. As was
noted also in \cite{ss}, the empty spaces in this table are not really
``empty'', but so far no application of them known to us has been
discussed in the literature.

\section{Conductance from symmetric spaces}
\label{sec-cond}
\setcounter{equation}{0}

Having identified the symmetric space to which the transfer matrix $M$
belongs, we can now solve the DMPK equation using the technique of
zonal spherical functions \cite{ss}.  In reference \cite{c1} the exact
procedure for solving the DMPK equation using the method of zonal
spherical functions was described in some detail. Here we will briefly
outline the steps, but the full details of the computation can be
found in the Appendix. A similar computation in the exactly solvable
$\beta=2$ case can be found in~\cite{br}.

The zonal spherical functions are eigenfunctions of the radial part of
the Laplace--Beltrami operator on a symmetric space:

\beq
\Delta_B'\phi_k(x)=\gamma_{\Delta_B'} (k)\phi_k(x)
\eeq

(where for brevity we have set $x=\{x_1,...,x_m\}$,
$k=\{k_1,...,k_m\}$ and $\gamma_{\Delta_B'} (k)$ are known eigenvalues
that may be functions of the roots, see for example \cite{OlshPere} or
\cite{ss}).  By using equation (\ref{eq:DMPKbis}) and the mapping
$\Delta_B'=J^{-1}BJ$ where $B$ is the DMPK operator given in
eq.~(\ref{defb}), we see that if $\phi_k(x)$ is a zonal spherical
function corresponding to the radial part of the Laplace--Beltrami
operator $\Delta_B'$ on a given symmetric space, then $J(x)\phi_k(x)$
is an eigenfunction of $B/(2\gamma)$ corresponding to the same
symmetric space (and with the same eigenvalue multiplied by
$1/(2\gamma )$).

The functions $\phi_k(x)$ have been studied by Harish--Chandra
\cite{HC}. They are related to irreducible representation functions of
groups and have a deep group theoretical significance \cite{ss,HC}.
They form a complete basis in the space of square--integrable functions
on the symmetric manifold, and can be used to express the analog of a
Fourier transform on the symmetric space.  Accordingly, the eigenvalue
density we are seeking can be written in the form

\beq
P(x,s)=J(x)f(x)=J(x)\int \tilde{f}(k) \phi_k(x) \frac{dk}{|c(k)|^2}
\eeq

where the integral over $k$ defines a Fourier transform on the
symmetric space of the function $\tilde{f}(k)$ describing the initial
conditions in $k$--space for the solution $P(x,s)$ of the DMPK
equation.  We choose $\tilde{f}(k_i)$ to be a gaussian:

\beq
\tilde{f}(k_i) \propto {\rm e}^{-\frac{k_i^2 s}{2\gamma }}
\eeq

where $s$ and $\gamma $ have the same meaning as in the DMPK equation.
This choice corresponds to ballistic initial conditions $l\gg L$
\cite{br}. The expression for $c(k)$ is
given by

\beq
\label{eq:cdef}
c(k)=\prod_{\alpha\in R^+} \frac{\Gamma(i(k,\alpha )/2)}{\Gamma (m_\alpha/2+i(k,\alpha )/2)}
\eeq

where the product goes over the positive roots of the root lattice
$R$, $\Gamma $ is the Euler gamma function, $m_\alpha $ the
multiplicity of the root $\alpha $, and $(k,\alpha )$ denotes a scalar
product.

The zonal spherical functions are known exactly for the ensembles of
transfer matrices corresponding to Dyson index $\beta =2$. For Dyson
indices $\beta =1,\ 4$ only asymptotic expressions are known.  We will
use the expression for $|c(k)|^2$ together with the asymptotically
known large--$x$ form of $\phi_k(x)$, valid for all values of $k$:

\beq
\phi_k(x)\sim J^{-1/2}(x) \sum_{r\in W} c(rk){\rm e}^{i(rk,x)}
\eeq

where $rk$ is the vector obtained by acting on $k$ with the element
$r$ of the Weyl group. We remind the reader that the Weyl group $W$ is the
symmetry group of the root system.  It consists in the most general
case of reflections and permutations of the root vectors.
 
We are particularly interested in the solution to the DMPK equation in
the insulating regime, where the peculiar feature of the carbon
nanotube -- the existence of one perfectly conducting channel in the
long--wire limit -- has the most relevant physical effect. In order to
keep our analysis as general as possible, we will discuss the solution
of the DMPK equation for a generic $BC_m$ root lattice (i.e. for a
root lattice of rank $m$ with general nonzero values for all three
types of roots). As before, we use the following notation for the root
multiplicities: $m_l=\eta $, $m_o=\beta $ and $m_s=\theta $.

Following the method outlined above, and described in full detail in
Appendix~\ref{app}, we obtain for the probability density $P(\{x_i\},s)$ of
the DMPK equation in the insulating regime

\begin{eqnarray}
\label{eq:P}
P(\{x_{n}\},s)&\propto &
\prod_{j>i} (\sinh^{2}x_{j}-\sinh^{2}x_{i})^{\, \beta /2}\, 
(x_{j}^{2}-x_{i}^{2}) \nonumber\\
&\times &\prod_{k}{\rm e}^{-\frac{x_k^2\gamma}{2s}}\,  
x_{k}^{h(\eta)+h(\theta)}\sinh^{\eta/2} (2x_{k}) \sinh^{\theta/2}  x_{k} 
\label{f1}
\end{eqnarray}

where $h(\eta)=0$  ($h(\theta)=0$) if no long (short) root is present in the 
lattice (i.e. if $\eta=0$ ($\theta=0$)), otherwise $h(\eta)=1$ ($h(\theta)=1$). 
In the present case $h(\eta)=h(\theta)=1$. 

After ordering the $x_{n}$'s from small to large and using the fact that in
this regime $1\ll x_{1}\ll x_{2}\ll\cdots\ll x_{m}$, we can approximate
the eigenvalue distribution as follows:

\begin{equation}
P(\{x_n\},s)\propto \prod_{i=1}^m{\rm e}^{-\frac{\gamma }{2s}(x_i-\bar{x}_i)^2}
\label{IR2}
\end{equation}

where 

\beq
\bar{x}_i=\frac{s}{\gamma}(\eta+\frac{\theta}{2}+\beta(i-1)) 
\eeq

In the strongly localized limit the conductance is dominated by the
first transmission eigenvalue. In this case, due to the
presence of the perfectly conducting channel, the first eigenvalue
actually gives the leading order correction to the contribution of
this channel.  We obtain the leading contribution $\delta g\equiv
\frac{G-G_0}{G_0}$ from equation (\ref{eq:g}). Using also 
(\ref{eq:Tlambda}) and (\ref{eq:subs}) we find:

\eq
\langle {\rm ln}\,{\delta g}\rangle \equiv -\frac{2L}{\xi}\approx 
{\rm ln}{T_1}\approx - 2\bar x_1=-\frac{2s(\eta+\frac{\theta}{2})}{\gamma}
\label{g1}
\en

where we have defined the localization length $\xi $ which we find to
be given by the expression

\eq
\label{eq:xi}
\xi=\frac{l\gamma }{\eta+\frac{\theta}{2}}
\en

The expression for $\gamma $ was given in equation~(\ref{eq:gamm}).
Substituting the values of the root multiplicities in the present case
($\eta=1,\ \beta=4,\ \theta=4$) and defining (following~\cite{T1})
$N=2m+1$ we find, in complete agreement with Takane's results,

\begin{equation}
\xi=\frac23(N-1)l
\end{equation}
\eqa
  - \langle \ln \, \delta g \rangle
      = \frac{3}{N-1} \left( \frac{L}{l} \right) 
             \\
         {\rm Var} (\ln \, \delta g)
      = \frac{2}{N-1} \left( \frac{L}{l} \right) 
\ena

where ${\rm Var}\, x \equiv \langle x^2 \rangle
-\langle x \rangle^2$ is the variance of $x$. It is interesting to
note that these two quantities can be combined into a universal ratio
which does not depend on the microscopic details of the model (i.e.,
on the value of $l$ or on the number of open channels) but only on
group theoretical quantities: the root multiplicities of the
appropriate symmetric space. Its general value is:

\eq
\label{eq:ur}
\frac{{\rm Var} (\ln \, \delta g)}{\langle \ln \, \delta g \rangle}=\frac{2}{\eta+\theta/2}
\en

This value only depends on the multiplicities of the short and long
roots and could play the same role in the insulating regime as the
universal conductance fluctuations play in the metallic regime (the
latter depend only on $\beta $, the multiplicity of the ordinary
roots).  In all the ``standard'' quantum wires this ratio is always 2,
because in the corresponding ensembles the root multiplicities have
the values $\eta=1$ and $\theta=0$. This makes the present carbon
nanotube case particularly interesting. Since in this case $\theta=4$
it is the first example\footnote{See~\cite{MC} for another possible
candidate, for which
there is, however, not yet any clear--cut physical evidence.} of a
nontrivial value of this universal ratio.

\section{Conclusion}

We have studied the electric conductance of single--layer carbon 
nanotubes with an odd number of conducting channels and belonging 
to the symplectic symmetry class, defined as the universality class 
of time--reversal invariant systems with strong spin--orbit scattering. 
The conductance of metallic carbon nanotubes can be studied within the 
Landauer formalism used for ordinary quantum wires. To obtain the
distribution of transmission eigenvalues needed to compute the 
conductance in this formalism, we have employed a technique related to
symmetric spaces.

After identifying the relevant symmetric coset space
$SO^*(4m+2)/[SU(2m+1)\times U(1)]$ to which the transfer matrix of the
system belongs, we show how knowledge of the theory of symmetric
spaces may be used to obtain a solution to the
Dorokhov--Mello--Pereyra--Kumar (DMPK) equation, a differential
equation describing the evolution of the transmission eigenvalues with
the length of the carbon nanotube. The solution is obtained by mapping
the DMPK equation onto the equation for free diffusion on the
symmetric space defined by the ensemble of transfer matrices. We then
find a suitable approximation to the distribution of eigenvalues in
the (would--be) insulating regime, where the existence of a conducting
channel even in the limit of an infinitely long nanotube in the
odd--channel case is known. Knowing the distribution of transmission
eigenvalues, we can compute the leading contribution to the physical
conductance of the carbon nanotube in terms of root multiplicities of
the symmetric space.  Our results can also be expressed in a form that
is in complete agreement with Takane's results.

In our case we have the maximal number of different types of roots
(i.e. a nonzero number of long, ordinary and short roots in the $BC_m$
root system corresponding to the symmetric space manifold). As a
consequence, we obtain the most general expression for the
localization length of such a system in terms of root multiplicities
(eq.~(\ref{eq:xi})). In addition we find the expression for the
universal ratio

\eq \nonumber
\frac{{\rm Var} (\ln \, \delta g)}{\langle \ln \, \delta g \rangle}=\frac{2}{\eta+\theta/2}
\en

in the insulating regime, which may play a role similar to that of
universal conductance fluctuations. These universal group theoretical
results are new and could only be derived using our technique based on
symmetric spaces.

\vskip1.0cm {\bf
Acknowledgments.} This work was partially supported by the
European Commission TMR programme HPRN-CT-2002-00325 (EUCLID).
U.M. was supported by the Lagrange foundation.

\appendix{}
\section{Appendix}
\label{app}
\setcounter{equation}{0}
\renewcommand{\theequation}{A.\arabic{equation}}

In this Appendix we give all the details of the calculation of
$P(\{x_i\},s)$. As we have already discussed, we can start from the
expression involving the Fourier transform

\beq
\label{eq:Fou}
P(x,s)=J(x)f(x)=J(x)\int \tilde{f}(k) \phi_k(x) \frac{dk}{|c(k)|^2}
\eeq

where $J(x)$ is the Jacobian on the symmetric space and

\beq
\tilde{f}(k_i) \propto {\rm e}^{-\frac{k_i^2 s}{2\gamma }}
\eeq

defines the initial condition. The function $c(k)$ is given by

\beq
\label{eq:cdef'}
c(k)=\prod_{\alpha\in R^+} \frac{\Gamma(i(k,\alpha )/2)}{\Gamma (m_\alpha/2+i(k,\alpha )/2)}
\eeq

Explicitly, for a root lattice with long, ordinary and short roots with
multiplicities $m_l=1,\ m_o=4,\ m_s=4$, $c(k)$ is given by 

\beq
c(k) = \prod_{i=1}^m \frac{\Gamma(ik_i)}{\Gamma(\frac{1}{2}+ik_i)}
\prod_{j>l}^m \frac{\Gamma(i\frac{k_j-k_l}{2})}{\Gamma(2+i\frac{k_j-k_l}{2})}
\frac{\Gamma(i\frac{k_j+k_l}{2})}{\Gamma(2+i\frac{k_j+k_l}{2})}
\prod_{k=1}^m \frac{\Gamma(i\frac{k_k}{2})}{\Gamma(2+i\frac{k_k}{2})}
\eeq

We will be interested in the insulating regime in which the variables
$\{k_i\}$ are small. In this regime we use the approximation 

\beq
\frac{\Gamma(iy)}{\Gamma (x+iy)}\sim \frac{i}{y}\ \ \ \ \ (y\to 0)
\eeq

Then $c(k)$ is approximated by

\beq
\label{eq:c-app}
c(k) \sim \prod_{i=1}^m k_i^{-2} \prod_{j>k}^m (k_j^2-k_k^2)^{-1}
\eeq

(where we have left out an irrelevant proportionality constant).
As was mentioned previously, the zonal spherical functions are not
exactly known in this case, but we can use the large--$x$ 
asymptotic formula

\beq
\label{phi}
\phi_k(x)\sim J^{-1/2}(x) \sum_{r\in W} c(rk)\, {\rm e}^{i(rk,x)}
\eeq

Here the sum is over all the configurations obtained by the action of
the elements of the Weyl group, i.e., the symmetry group of the
restricted root system. The elements of the Weyl group are
reflections and permutations. Since $c(k)$ is invariant under
reflections, only the even part $\sum_r c(rk)\, {\rm cos}(rk,x)$ of
$\sum_r c(rk)\, {\rm e}^{i(rk,x)}$ survives. Under permutations,

\beq
\label{eq:perm}
\sum_{r=p} c(rk)\, {\rm cos}(rk,x) = \sum_{r=p} (-1)^{\sigma(p)} c(k)
\, {\rm cos}(k_{p(j)}x_j) \sim  
\frac {\sum_p (-1)^{\sigma(p)}{\rm cos}(k_{p(j)}x_j)}
{\prod_{i=1}^m k_i^2 \prod_{j>k}^m (k_j^2-k_k^2)}
\eeq

Defining

\beq
{\rm det}_{\scriptstyle 1\leq i,j \leq m}[C_{ij}]
\equiv \sum_p (-1)^{\sigma(p)}{\rm cos}(k_{p(j)}x_j)
\eeq
\beq
{\rm det}_{\scriptstyle 1\leq k,l \leq m}[\Delta_{kl}]\equiv
\prod_{k>l}^m (k_k^2-k_l^2)= 
{\rm det}_{\scriptstyle 1\leq k,l \leq m} k_l^{2(k-1)}
\eeq

where ${\rm det}\Delta$ is a Vandermonde determinant, we finally
obtain from eqs.~(\ref{eq:Fou}), (\ref{eq:c-app}), (\ref{phi}) 
and (\ref{eq:perm})

$$
P(x,s)\propto \sqrt{J(x)} \int \prod_{n=1}^m \left( {\rm d}k_nk_n^2
{\rm e}^{-\frac{k_n^2s}{2\gamma}}\right) {\rm det}\Delta \, {\rm det} C
$$

$$
=\sqrt{J(x)}\int \prod_{n=1}^m {\rm d}k_n\,  
{\rm det}_{1\leq i,j \leq m}
\left[ \sum_{k=1}^m k_k^{2i}{\rm e}^{-\frac{k_k^2s}{2\gamma}}{\rm cos}(k_kx_j) \right]
$$

\bea
=\sqrt{J(x)} \sum_\sigma \sum_{k=1}^m ...\sum_{k'=1}^m (-1)^{\sigma(p)}
\int {\rm d}k_k k_k^2 {\rm e}^{-\frac{k_k^2s}{2\gamma}}{\rm cos}(k_kx_\alpha ) ...
\int {\rm d}k_{k'} k_{k'}^{2m} {\rm e}^{-\frac{k_{k'}^2s}{2\gamma}}{\rm cos}(k_{k'}x_\nu ) 
\nonumber \\ 
\eea

where we have used ${\rm det}A{\rm det}B={\rm det}[AB]$ and expanded
the determinant according to ${\rm det}A=\sum_\sigma
(-1)^{\sigma(p)}A_{1\alpha}A_{2\beta}...A_{m\nu}$.  We now use the
known result for the integral

\beq
\int_0^\infty {\rm d}k k^{2n}{\rm e}^{-b^2k^2}{\rm cos}(ak)\propto
{\rm e}^{-\frac{a^2}{4b^2}}H_{2n}\left(\frac{a}{2b}\right)
\eeq

where $H_{2n}$ denotes a Hermite polynomial of order $2n$, to obtain

\beq
P(x,s)\propto \sqrt{J(x)} \sum_\sigma (-1)^{\sigma(p)} \sum_{k=1}^m ... \sum_{k'=1}^m
{\rm e}^{-\frac{x_\alpha^2\gamma}{2s}}H_2\left(\frac{x_\alpha}{\sqrt{\frac{2s}{\gamma}}}\right)
... {\rm e}^{-\frac{x_\nu^2\gamma}{2s}}H_{2m}\left(\frac{x_\nu}{\sqrt{\frac{2s}{\gamma}}}\right)
\eeq

Setting 

\beq 
\frac{x_\alpha}{\sqrt{\frac{2s}{\gamma}}}\equiv \tilde{x}_\alpha
\eeq 

for brevity, the expression 

\beq
\sum_\sigma (-1)^{\sigma(p)} H_2(\tilde{x}_\alpha) ... H_{2m}(\tilde{x}_\nu)
\eeq 

is a determinant of Hermite polynomials of even order.
The general form of $H_{2n}(x)$ is 

\beq
H_{2n}(x)=c_0(x^{2n}+c_1x^{2n-2}+...+c_n)
\eeq

where $c_n\neq 0$. To evaluate the determinant, we take out the
irrelevant constants $c_0$ in front of the Hermite polynomials from
the determinant so that we have monic polynomials in the latter. By
adding an appropriate multiple of the first row to each of the other
rows, we can eliminate the constants $c_n$ from each matrix element,
except for the ones in the first row.  Using that
$H_2(x)=4(x^2-\frac12)$ the determinant then takes the form

$$
{\rm det}_{1\leq i,j \leq m} [H_{2i}(\tilde{x}_j)]\propto \left|
\begin{array}{cccc}\tilde{x}_1^2-\frac12 & \tilde{x}_2^2-\frac12 & ... 
 & \tilde{x}_m^2-\frac12\\
\\
\tilde{x}_1^2 p_2(\tilde{x}_1) & \tilde{x}_2^2 p_2(\tilde{x}_2) & ...
 & \tilde{x}_m^2 p_2(\tilde{x}_m)\\
... & & & \\
\tilde{x}_1^2 p_{2(m-1)}(\tilde{x}_1) & \tilde{x}_2^2 p_{2(m-1)}(\tilde{x}_2) 
& ... & \tilde{x}_m^2 p_{2(m-1)}(\tilde{x}_m)\end{array}\right|
$$

\bea
=\tilde{x}_1^2...\tilde{x}_m^2 \left| \begin{array}{cccc}
1 & 1 & ... & 1 \\
\\
p_2(\tilde{x}_1) & p_2(\tilde{x}_2) & ... & p_2(\tilde{x}_m)\\
... & & & \\
p_{2(m-1)}(\tilde{x}_1) &  p_{2(m-1)}(\tilde{x}_2) & ... &  
p_{2(m-1)}(\tilde{x}_m)\end{array}\right| + \nonumber \\  
\nonumber \\
\left| \begin{array}{cccc}-\frac12 & -\frac12 & ... & -\frac12\\
\\
\tilde{x}_1^2 p_2(\tilde{x}_1) & \tilde{x}_2^2 p_2(\tilde{x}_2) & ...
 & \tilde{x}_m^2 p_2(\tilde{x}_m)\\
... & & & \\
\tilde{x}_1^2 p_{2(m-1)}(\tilde{x}_1) & \tilde{x}_2^2 p_{2(m-1)}(\tilde{x}_2) 
& ... & \tilde{x}_m^2 p_{2(m-1)}(\tilde{x}_m)\end{array}\right|
\eea

where $p_{2n}$ denotes a polynomial of degree $2n$ and we have used
general properties of determinants.  Since we are interested in the
insulating regime where $1\ll x_1 \ll x_2 \ll ... \ll x_m$, we can
neglect the second determinant in the last line of the equation, whereas the
first determinant can be re--written as a Vandermonde
determinant by simply writing each line of the determinant as a linear
combination of the other rows.  Then we obtain

$$
P(\{x_i\},s) \propto \sqrt{J(x)} \prod_{i=1}^m {\rm e}^{-\frac{x_i^2\gamma}{2s}} 
{\rm det}_{1\leq i,j \leq m}
\left[H_{2i}\left(\frac{x_j}{\sqrt{\frac{2s}{\gamma}}}\right)\right]\nonumber 
$$
\bea
\label{eq:fP}
=\prod_{i=1}^m {\rm sinh}^{\eta/2}(2x_i) \prod_{j>k}^m {\rm sinh}^{\beta/2}(x_j-x_k)
{\rm sinh}^{\beta/2}(x_j+x_k) \prod_{l=1}^m {\rm sinh}^{\theta/2} x_l 
{\rm e}^{-\frac{x_l^2\gamma}{2s}} \prod_{p>q}^m (x_p^2-x_q^2) \prod_{r=1}^m x_r^2
\nonumber \\
\eea

In the last line we have used (\ref{eq:jacobianox}) with $m_l,\ m_o,\
m_s$ instead of the explicit root multiplicities. After using the identity for
hyperbolic functions

\beq
{\rm sinh}(x_j-x_k){\rm sinh}(x_j+x_k)= {\rm sinh}^2x_j-{\rm sinh}^2x_k
\eeq

(\ref{eq:fP}) becomes the expression reported in eq.~(\ref{eq:P}).

\end{document}